# A Global View of Cognitive Structure and Implications for Instruction and Assessment

Andy Terrel[a], Computer Science Dept., University of Chicago, 1100 E. 58th St., Chicago, IL 60637 and Beth Thacker, Physics Department, MS 41051, Texas Tech University, Lubbock, TX 79409-1051

**Abstract**

We assert that models of cognitive structure all have the same basic features. These basic features, independent of the model, have important implications for instruction and assessment. We describe the basic features of models of cognitive structure, giving examples from our research. We then discuss the implications for instruction and assessment.

[a] At the time of this work, Andy Terrel was a student in the Physics Department, MS 41051, Texas Tech University, Lubbock, TX 79409-1051.

**A Global View of Cognitive Structure and Implications for Instruction and Assessment**

Andy Terrel[a], Computer Science Dept., University of Chicago, 1100 E. 58th St., Chicago, IL 60637 and Beth Thacker, Physics Department, MS 41051, Texas Tech University, Lubbock, TX 79409-1051 \

**I. INTRODUCTION**

Recent literature has pointed out both the implications of cognitive studies for teaching physics and the need for a theoretical framework for physics education research, possibly most comprehensively, in Refs. 1 and 2. Students' reasoning processes, and cognitive structure, in general, are very complex and have been the subject of many papers[1-2, 3-14]. As in the study of any complex system, it is useful to identify simple, underlying features, apparent in all systems. These features constitute a very general model, which can be refined and expanded upon as more detailed features are identified. We believe that the underlying features, the coarse structure, of models of cognition have important implications for instruction and assessment. In this paper, we discuss these features and give examples from our research.

Historically, the focus of physics education research has been on the study of conceptions.[3] Conceptions have been studied without being very well defined. They are generally implicitly considered to be large knowledge structures, which may or may not be robust. Often the focus is on the difference in novice and expert conceptions. The implications for instruction are that the "incorrect" conceptions of novices can be identified through research and that instructional materials can be developed which confront students' incorrect conceptions, leading them to alter their conceptions and to develop models of physical phenomena consistent with scientists' models.

More recently, researchers have focused on more detailed models of the structure and dynamics of students' reasoning. The focus has shifted to a study of smaller knowledge structures of which conceptions are composed and the dynamics of creating larger knowledge structures from smaller ones. There are different models of the process of conceptual change evolving, including the details of how small knowledge structures are linked, how links are created and destroyed and the role of external stimuli in constructing larger knowledge structures.[2, 4-13]

These more detailed models have similar underlying features. We believe that these features have important implications for instruction and assessment. There are a number of reasons that these underlying features of cognitive models have not been recognized as a general model of cognition. Some of these are 1) researchers have not used a common terminology in the labeling of elements of cognitive structure and 2) they believe that more detail, particularly about the dynamics of students' reasoning, is needed for the model to have clear implications for teaching and learning.

We discuss the underlying features of cognitive structure, giving examples from our research. We, then, discuss implications of these features for instruction and assessment.

## II. RESEARCH

We report results from our research, simply because they have not been reported elsewhere. We are not presenting a new model of cognitive structure based on our research. We are presenting examples that illustrate the basic, underlying features of all models of cognitive structure. It is simply easier, for us, to give examples from our own research, rather than someone else's.

The goal of our research, and the goal of this paper, is to present examples that illustrate the basic, underlying features of cognitive structure and the implications of these features for instruction and assessment.

### A. The Students

We analyzed interviews of students on two different topics: 1) modern physics and 2) fluid dynamics. In each case we analyzed pre-interviews that had been done for other studies and re-analyzed the data, focusing on evidence of general, underlying features of cognitive structure.

Since we were only interested in identifying general features of cognitive structure, not in comparing students' answers to correct answers or comparing answers of students taught by different instructional methods, we do not present details about the students or the interviews. These are presented elsewhere.[14-15] We only briefly describe the context of the interviews on each topic, mostly so the reader can understand the examples later in our discussion.

We chose to analyze pre-interviews, and in particular, pre-interviews of students who had had some formal physics instruction, but not formal instruction on the topics on which they were being interviewed. The reason for this is that it is in these interviews that we can watch students develop models. They have not had formal instruction in the scientific models accepted to describe the phenomena on which they are being questioned, but they have had some physics instruction and this is an ideal place to watch students develop models. In short, an ideal place to observe the dynamics of cognitive structure.

**i. Modern physics interviews**

Students in a modern physics class were interviewed individually before instruction on the photoelectric effect. They were shown an apparatus that consisted of a carbon arc lamp and an electroscope with a piece of zinc attached to the top. The students were told that the carbon arc lamp emitted both UV and visible radiation. Three cases were demonstrated:
> 1) The electroscope was charged negatively and then light from the carbon arc lamp was shone on it. This discharged the electroscope.
> 2) The electroscope was charged positively and then light from the carbon arc lamp was shone on it. There was no significant change in the electroscope.
> 3) The electroscope was charged negatively and a piece of glass was placed between the arc lamp and the electroscope when the light was shone on it. There was no significant change in the electroscope.

In the interviews, the students were first asked to describe what happened and then to explain why. Questions were also asked to probe the students' understanding of the concepts of light, intensity, energy and frequency as related to the photoelectric effect. The details of this research have been reported elsewhere.[14]

**ii. Fluid dynamics interviews**

Students in an introductory algebra-based physics class taught a laboratory-based curriculum designed specifically for students in the algebra-based course,[16] participated in a half-hour pre-interview before instruction in fluid dynamics. The students had already covered mechanics (including topics such as kinematics, dynamics, energy and momentum, etc.) at the time of the pre-interviews. The students were asked questions about systems of pumps and pipes. They were asked to rank the flow rate, velocity and pressure of the fluid at different points in the systems, with the pump turned on and the system filled with water. They were asked questions about the reasoning behind their choices. The details of this research will be reported elsewhere.[15]

**B. Examples of the Underlying Features of Cognitive Structure from our Research**

When students are asked to answer a question about an observable phenomenon, the structure of their answers reflects a cognitive model that they are constructing to explain the phenomenon.

We present examples of the underlying features of cognitive structure from our interviews. We, as other researchers[4-13], found evidence for small knowledge structures that appeared to be irreducible to the person being interviewed. These were knowledge elements used by the person to form larger knowledge structures, but whose composition of even smaller knowledge structures was not acknowledged by or not relevant to that person. We labeled these fundamental reasoning elements. In relation to other existing labeling schemes, fundamental reasoning elements are a type of resource[6], which includes, but is broader than, the sub-category phenomenological primitives[7] (p-prims).

Again, we do not wish to focus primarily on the labeling scheme, but on the general structure of the features.

We also found evidence of groupings of fundamental reasoning elements and the connections between them. These we labeled patterns of association, to be consistent with the other literature. We describe these below.

**i. Fundamental Reasoning Elements**

A fundamental reasoning element is a small knowledge structure that appears to be irreducible to the person being interviewed. In everyday language, a simple fundamental reasoning element is "a fact," something not to be doubted, questioned or analyzed, a starting point for answering a question or developing a model. The students interviewed used these statements as known "facts" that they did not have to justify or defend.

We classified different kinds of fundamental reasoning elements that we observed in our interviews. Some examples are:

a) Fundamental reasoning elements originating from a source which holds some type of authority in the field, such as an instructor or a book. For example:

"Protons are positively charged."

b) Fundamental reasoning elements originating from commonly made statements. For example:

"…it acts like both a particle and a wave."

The student making this statement applies it as a fact, without structure.

c) Fundamental reasoning elements originating from abstraction or generalization of some process. For example:

"Some is leaving it and some is going on to it, kinda equilibrium and maybe like a reaction."

"Because it wants to go towards equilibrium, right? It is like putting food color in a glass, you will see it but put it in a pool it will disperse."

Both of these statements indicate the generalization that "things come to equilibrium." This type of fundamental reasoning element is most obviously a phenomenological primitive (p-prim).

d) Fundamental reasoning elements originating from observation. For example:

"I saw two carbon rods that moved together to produce an arc."

"The electroscope was charged positively,…, then light from the carbon arc was shown on it."

These are statements of what the person observed.

**ii. Patterns of Association**

Patterns of association are groupings of fundamental reasoning elements and the connections between them. Patterns of association may be simple or complex. A simple pattern of association is created by choosing two or more fundamental reasoning elements that may be related to each other based on basic features; it is simply a consciousness that the fundamental reasoning elements may be related. A complex pattern of association is created by choosing two or more fundamental reasoning elements based on logic and motivation, reflecting on them, and relating them to create a solid connection.

An example of a simple pattern of association is shown in Fig. 1. The two fundamental reasoning elements "There must be a positively charged particle" and "Would it be protons?" are grouped together into a pattern of association. It is simply a consciousness that they may be related.

An example of a complex pattern of association is shown in Fig. 2. Four fundamental reasoning elements are joined to form a pattern of association. Fig. 3 shows another way a complex pattern of association can be created. In Fig. 3, part of one pattern of association is combined with other fundamental reasoning elements to create a new pattern of association. The patterns have been created because they are consistent and coherent to the individual creating the pattern. However, they are not consistent with scientists' models.

Patterns of association can be more complex than those illustrated in Figs. 2 and 3. Once a pattern of association is formed, it can be used in the same way as a fundamental reasoning element in the formation of more complex patterns of association. An example of a more complex pattern of association is illustrated in Fig. 4. A pattern of association is used as a fundamental reasoning element in the formation of a new pattern of association.

It is also possible to hold contradictory patterns of association, if there is no motivation, logic or experimental evidence available to force a choice of one over the other. The best example of this came from our interviews on fluid dynamics. Consider the following extract from an interview with a student discussing the speed of the fluid through a narrow section of pipe in a closed fluid system:

Interviewer: Why would it, when you put this constriction here, would it cause it to speed up? What's causing it to speed up?

Student: OK. I'm reversing my train of thought. I think it actually slows because, going back to…I would have to think of something…thinking of blood vessels: open it allows more blood to flow through a certain area. So when they're restricted, less blood is able to flow for the rate of time. So thinking about that, I would think that it does slow. The rate of flow is slower than it is through A and C. Just based on constricting…The reason I was thinking faster is I was thinking moving all this water it has nowhere else to go – it has to go through there. That same amount of water that's in it would have to go through here, but in less space, so that would cause it to go faster. But I think I was wrong, because if you constrict it, it will slow it down. Not as much water is able to go through.

Later in the same interview said:

Student: Maybe the velocity would just remain the same and pressure is the only thing that changes.

The student was holding three contradictory patterns of association (the velocity decreased, the velocity increased and the velocity stayed the same), as illustrated in Fig. 5, at the same time. Each pattern of association had a structure, but there was no experimental evidence or logic to choose one above the other at that point in time.

In summary, students create patterns of association by choosing and grouping fundamental reasoning elements, using logic and motivation, reflecting on them and relating them to create a connection. We, as other researchers, also made the observations that patterns of association 1) are dynamic, always changing, 2) may be tightly (very stable) or loosely (easily disassociated) bound. There are many ways to create a complex pattern of association. We illustrate and discuss this below.

## IV. A GENERAL SUMMARY OF THE UNDERLYING FEATURES OF COGNITIVE STRUCTURE

The general structure of patterns of association, from our research and that of others[2,4-13], can be illustrated as in Fig. 6. This figure is meant to illustrate the general structure of patterns of association, not any detail. Whether different researchers have labeled different elements resources, or facets or phenomenological primitives, is irrelevant – the general structure is the same. In our terminology, fundamental reasoning elements are combined to form patterns of association and, once formed, patterns of association can be combined to form more complex patterns of association. We have indicated more complex patterns of association by shapes of different colors and at different levels.

If a person is motivated to disassemble a pattern of association, it may affect many other patterns of association. We illustrate this in Fig. 7. A person has chosen to disassemble a pattern of association at level 2, which causes many patterns of association at higher levels to disassemble, changing the complete structure of that person's patterns of association.

We, as other researchers (see, for example Ref. 2), also identified some general features of patterns of association:

1) The development of patterns of association is an individual's attempt to create consistent and coherent cognitive structures based on logic, motivation and external evidence. Patterns of association are created to be consistent and coherent with other patterns of association and with external evidence, but it is an internal consistency. The patterns are consistent to that individual.
2) This is a dynamic process. Patterns of association are always changing. However, patterns of association may remain stable for a long time, if there is no motivation, logic or experimental evidence which causes the person to re-analyze the patterns of association and decide to disassemble them.
3) Patterns of association may be stable (tightly bound) or unstable (easily disassociated). If it is loosely bound, it is consistent and coherent with external information *and* with that person's other patterns of association at the present time, but it is easily disassociated into its parts. In everyday language, the person is not very sure of it. A tightly bound pattern of association is also consistent and coherent, but it is not easily disassembled into parts. In everyday language, the person is very sure of it (which again does not mean that it is "correct" in the sense of agreeing with models that scientists have developed).
4) A pattern of association is not disassembled unless the person perceives an inconsistency with other patterns of association or external stimuli.
5) It is possible to hold contradictory patterns of association, if there is no motivation, logic or experimental evidence available to force a choice of one over the other.

While these have been studied in more detail and listed using different terminology, we believe that these basic features and the general structure of patterns of association as illustrated in Figs. 6 and 7, have important implications for teaching and assessment.

**IV. IMPLICATIONS FOR INSTRUCTION AND ASSESSMENT**

We assert that one of the primary goals of a physics course is that students have the ability to assemble patterns of association that are both internally consistent and coherent *and* consistent and coherent with experimental evidence and existing scientific models.

If we accept this as a goal of the course, we need to develop instructional strategies that will help us achieve that goal and have assessment that will indicate whether we have been successful. This is a very different goal from "learning content." It has to do with "letting the students do the thinking," helping the students to learn to develop consistent and coherent cognitive structures that are also consistent with experiment. In short, we now have the "experimental evidence," the underlying features, or coarse model of cognitive structure (based on experiment), that supports alternative methods of instruction and does not support traditional methods. Because patterns of association are constructed by an individual, to be consistent to that individual, the focus of teaching must be on helping the student construct consistent and coherent patterns, with the instructor as guide. This supports the gut feeling of many teachers that inquiry-based methods[17] are more effective at teaching students to develop consistent and coherent patterns of association than traditional methods and that the effect of lecturing is to demonstrate that the lecturer has developed consistent and coherent patterns of association, but not the student.

This may seem like a very broad claim to many teachers, who are used to research being expressed as a very detailed model, but we claim that it is the common, underlying features of all models of cognitive structure that have implications for instruction and assessment. More detailed models may lead to more detailed instructional materials or assessments, but we claim that it is not the detail, but the general features that support some teaching methods over others and have implications for assessment.

**A. Instruction**

Based on the general features of patterns of association, an important aspect of the learning process is for the student to learn to reflect on and relate cognitive structures until they have a cognitive structure that is coherent and consistent with other cognitive structures *and* with experimental evidence and existing scientific models. An important part of instruction, then, is to help the student do this.

Some aspects of instruction support this goal more actively than others. We name some aspects of instruction (the list is not inclusive) that actively promote the reflection and relation of cognitive structures below:

1) **Pre-tests:** Pretests help instructors identify stable cognitive structures that may be inconsistent with scientists' models. They also promote reflection and relation of knowledge structures by the student and can promote motivation (a recognition that some cognitive structures may have to be restructured).

2) **Instructional methods that promote the teacher as a guide:** Instructional methods such as inquiry-based instruction, discovery methods, instruction that uses Socratic dialog, etc. that promote the instructor as a guide, focus explicitly on helping students to develop cognitive structures that are consistent, coherent and consistent with experimental evidence.

3) **Laboratory-based instructional methods:** Laboratory-based instructional methods, such as "Workshop Physics,"[18] or laboratories in which the student is responsible for independently (or in a group) setting up equipment, taking and analyzing data, developing models based on the data, etc., as opposed to "cookbook labs" or reading about the results of experiments, require the student to reflect on and relate cognitive structures *and* develop patterns of association consistent with experimental evidence.

4) **Interactive engagement methods:** Any method that allows the student time to reflect and construct his/her own patterns of association, such as "Think, Pair Share,"[19] "Peer Instruction,"[20] etc. during the course of instruction.

5) **An explicit focus on problem solving strategies:** Any method that focuses on problem solving strategies, such as Cooperative Group Problem Solving.[21]

6) **Problem solving which requires the student to explain their reasoning:** Whether students are required to explain their reasoning to other students during a collaborative problem solving session, or to explain their reasoning in writing on homework or quiz or exam problems, the process of explaining, requires the construction of coherent and consistent cognitive structures.

Again, the list is not inclusive. These aspects of instruction focus explicitly on helping the student to reflect on and relate cognitive structures. In everyday language, these aspects of instruction help the "students do the thinking". We expect that if you employ one of these aspects of instruction in your course, you will be more successful in achieving the stated goal, because these aspects of instruction focus on helping students assemble consistent and coherent patterns of association. We will discuss the assessment of whether this goal has been achieved in the next section.

We could equally well make a list of aspects of instruction that do not explicitly promote reflection and relation of cognitive structures. This list would include aspects such as: 1) traditional instruction (lecture), because the student is not explicitly encouraged to reflect on and relate cognitive structures (the student may do it or may not), 2) reading a text, for the same reason, and 3) homework and exam problems that are only multiple choice, because it is possible to get the correct answer without developing cognitive structures that are consistent and coherent.

**B. Assessment**

If we take the above stated goal seriously, not only do we need to use instructional strategies that promote reflection and relation of cognitive structures, but assess whether we have achieved our goal of helping students develop consistent and coherent cognitive structures. Traditionally in physics, we have done this by asking written questions that require students to show their calculations. More recently there has been a focus by some instructors on requiring students to explain their reasoning, which is somewhat different, as it includes verbal explanations in addition to mathematical calculations. This type of assessment provides evidence of the students' ability to develop consistent and coherent cognitive structures.

Recently, there have been movements away from written explanations as assessment instruments. Most often, this is a movement towards multiple choice assessment, and a movement driven by economic considerations. In our opinion, these are movements away from assessment that is effective at assessing students' ability to create coherent and consistent cognitive structures, focusing only on whether or not the student has developed cognitive structures that can predict the same results as existing scientific models.

We examine three methods of assessment and the effectiveness of each method at assessing students' cognitive structures:

1) **Interviewing students:** Students are interviewed and the transcripts of the interviews are analyzed, looking for cognitive structure that is internally consistent and coherent and consistent with experimental evidence.

2) **Written questionnaires that include explain your reasoning:** Students are required to show their work and explain their reasoning in answering written homework, quiz or exam problems. The solutions provide evidence of consistent and coherent cognitive structures.

3) **Multiple choice questionnaires:** Only multiple choice questionnaires carefully constructed based on research can, to a certain extent, provide evidence of consistent and coherent cognitive structures.

We address methods (1) and (3) first. When students are interviewed the interviewer gains a wealth of information on cognitive structure. It is from this method of assessment that 1) researchers can develop models of cognitive structure and 2) instructors and researchers can very effectively assess the ability of students to develop coherent and consistent cognitive structures. This method of assessment is ideal for a researcher, but much too time consuming for most instructors to use in each of their classes.

Multiple choice questionnaires yield very little information on the ability of students to develop coherent and consistent cognitive structures. At best, these assessment instruments can be used to indicate whether students have developed cognitive structures that predict results consistent with existing scientific models or with common "incorrect" conceptions, but with very little information on the consistency and coherency of the structure. By requiring students to explain their reasoning to a multiple choice question, one finds that the explanations are often inconsistent with the answer and many students (this is often as high as 30 – 50% of the students who have chosen the correct answer) cannot correctly explain their choice of the right answer.[22-23]

We would like to make an analogy with the present day food industry in the United States. Multiple choice exams are like eating at MacDonalds®, as opposed to the organic food market (interviewing students) down the street. While we are aware of the economic reasons for "eating at MacDonalds®", we believe there is a point at which it is important to make a "healthy" choice, even if it costs a little bit more. We believe it is important to use assessment instruments that assess the ability of students to develop consistent and coherent cognitive structures and we make an argument for "Subway®" assessment instruments -- written questionnaires that require students to explain their reasoning -- as a somewhat more expensive, but much more effective alternative to multiple choice exams.

Many non-traditional methods of instruction, such as the aspects of instruction listed above, focus explicitly on the goal of helping students learn to assemble patterns of association that are both internally consistent and coherent *and* consistent and coherent with experimental evidence and existing scientific models. Yet, when comparing them to traditional methods of instruction, we often don't assess whether this goal has been achieved.

We claim that "MacDonalds®" assessment (multiple choice) is not sufficient for determining whether students have achieved the stated goal, even those assessments that have been developed to investigate whether students have some common "incorrect" conceptions. Included in this type of assessment are standard physics assessment instruments, such as the Force Concept Inventory (FCI)[24] or the Force and Motion Conceptual Evaluation (FMCE)[25] and others.[26] It is not that these assessments don't serve a purpose, but they do not provide evidence of whether students have developed consistent and coherent cognitive structures. More appropriate would be "Subway®" type assessment instruments, which give much more detail about students' development of consistent and coherent cognitive structures. We encourage the physics education research community to develop a battery of "Subway®" assessment instruments that can be used by the average instructor to assess cognitive skills. These assessment instruments would not be multiple choice, but would require students to explain their reasoning. While individual instructors and individual physics education research groups[27] have used written questionnaires that require students to explain their reasoning, there is not a set(s) of Subway® type assessment instruments, including an appropriate rubric, which could be used across different universities to assess students' ability to develop coherent and consistent cognitive structures.

We are also concerned about the use of only multiple choice exams in other contexts, such as the physics component of standardized assessments, from those used in statewide assessments at all levels K-12 to college, graduate and professional school entrance exams, but we will address these in another paper.

## V. CONCLUSIONS

The basic, underlying features of models of cognitive structure have important implications for instruction and assessment. Based on general statements about cognitive structure, we could say that an important aspect of the learning process is for students' to learn to reflect on and relate cognitive structures until they have a consistent and coherent cognitive structure that is consistent with experimental evidence and existing scientific models. If one of the primary goals of a physics course is that students have the ability to assemble consistent and coherent patterns of association, then aspects of instruction that focus explicitly on helping the student to reflect on and relate cognitive structures will be more successful at achieving the stated goal. We need to choose and develop instructional strategies that will help us achieve this goal and employ assessment that will indicate whether we have been successful.

**Example of a Simple Pattern of Association**

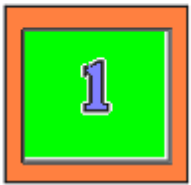

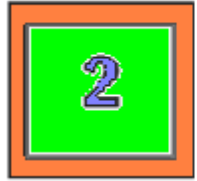

**Fundamental Reasoning Elements:**

1) There must be a positively charged particle.

2) Would it be protons?

**Figure 1a**

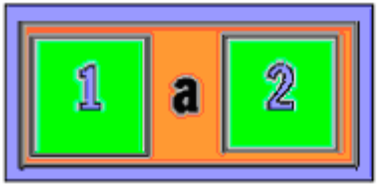

The two Fundamental Reasoning Elements are grouped together. The Simple Pattern of Association is a consciousness that two Fundamental Reasoning Elements are related.

**Figure 1b**

## Example of a Complex Pattern of Association

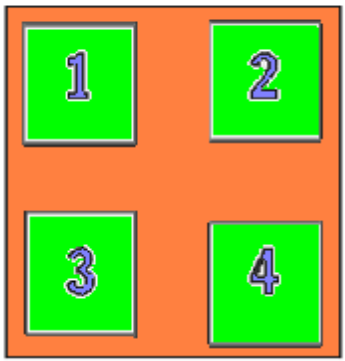

**Fundamental Reasoning Elements:**
1) The electroscope is negatively charged.
2) Negative charge is discharged by the positive electric field.
3) Protons are positively charged.
4) The lamp is the only stimulus.

**Figure 2a**

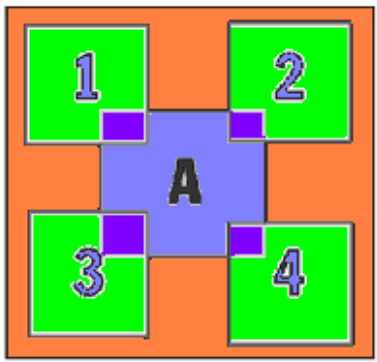

The Fundamental Reasoning Elements are combined to form a Complex Pattern of Association:

A. A positive electric field, which may be caused by protons, comes from the lamp to discharge the electroscope.

**Figure 2b**

## An Example of a Complex Pattern of Association

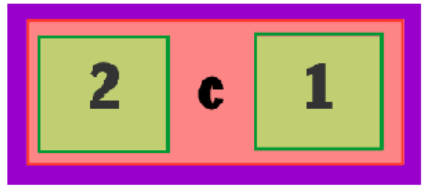

A simple pattern of association consisting of the Fundamental Reasoning Elements:
1) In a vacuum, the electroscope becomes more positive and always increases in charge.
2) In a vacuum, the electroscope will not lose any charge due to air.

**Fig. 3a**

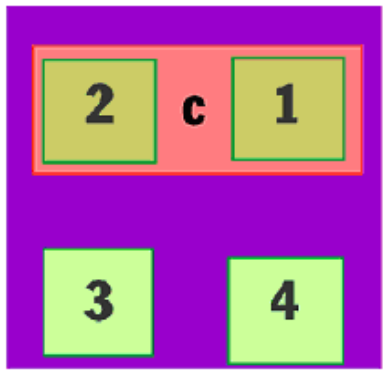

Three Fundamental reasoning elements are chosen to create a new reasoning element. In the process, a simple pattern of association is disassembled.

Fundamental Reasoning Elements chosen:

2) In a vacuum, the electroscope will not lose any charge due to air.
3) There is no air in a vacuum.
4) According to my theory, the positive charge has to travel through the air.

**Fig. 3b**

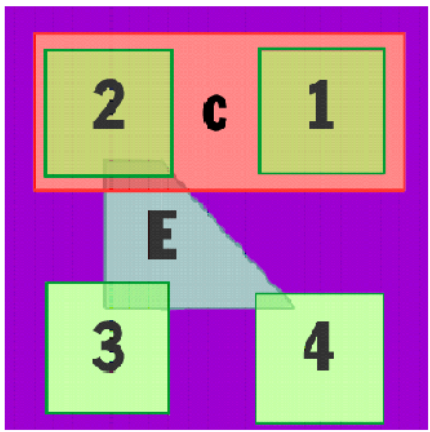

The Fundamental Reasoning Elements are combined to form a pattern of association:

The electroscope will not become more positively charged in a vacuum because the positive charge [from the carbon arc lamp] has no place to flow.

**Fig. 3c**

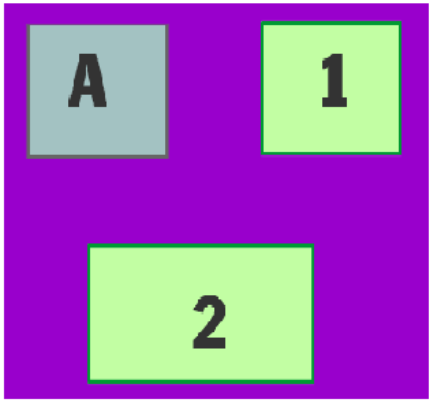

A pattern of association A (Fig. 2) is used as a Fundamental Reasoning Element.

Three fundamental reasoning elements are chosen:
A) A positive electric field, which may be caused by protons, comes from the lamp to discharge the electroscope.
1) The lamp is the only stimulus.
2) A positive electric field is formed by positively charged particles.

**Fig. 4a**

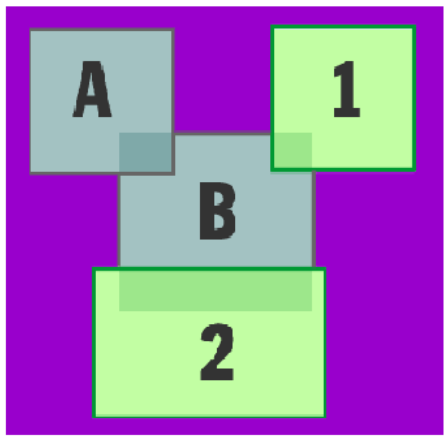

A new pattern of association is created in order to more fully describe A:

Some positively charged particles from the lamp discharge the electroscope.

**Fig. 4b**

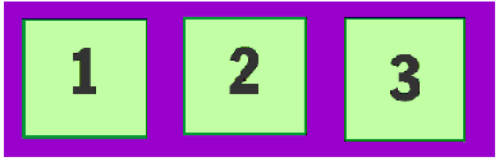

A student holds three contradictory patterns of association at the same time:

1) The velocity increases.

2) The velocity decreases.

3) The velocity remains the same.

**Fig. 5**

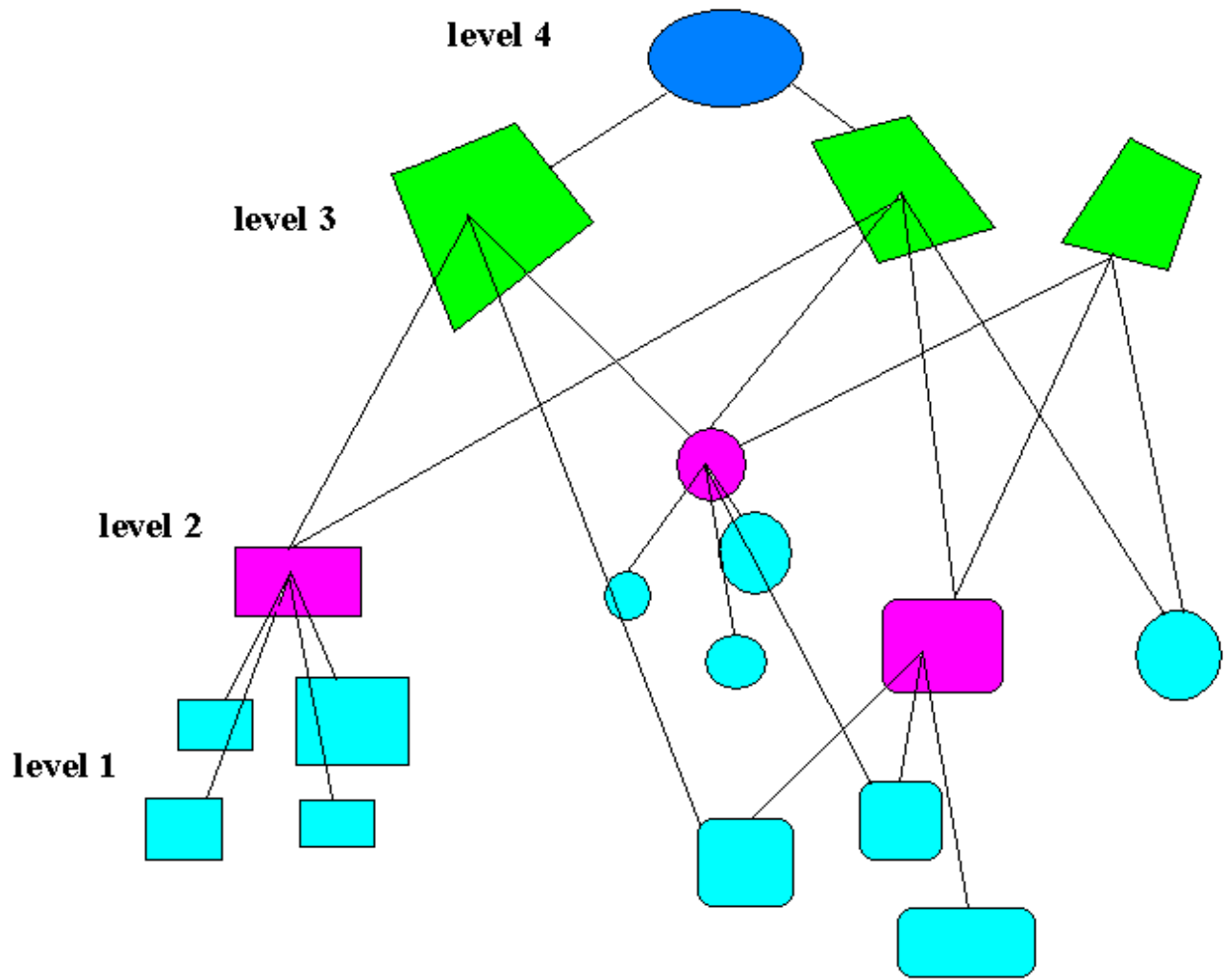

**Figure 6**

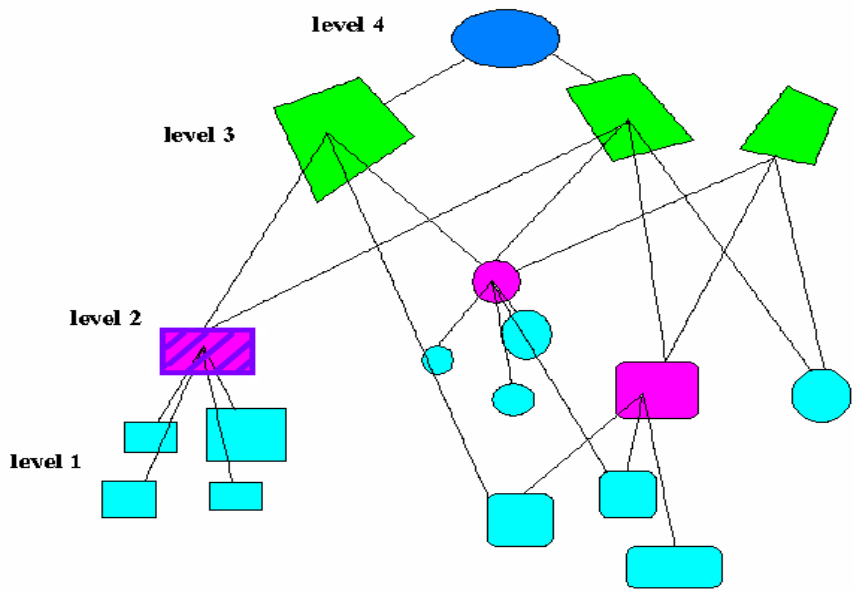

**Fig. 7a**

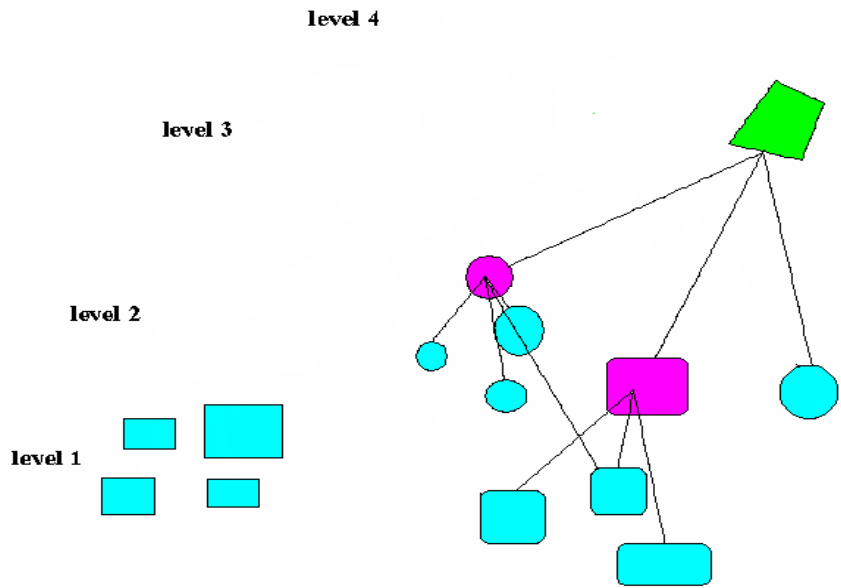

**Fig. 7b**

**Figure Captions**

Figure 1: Figure 1a shows the existence of two fundamental reasoning elements and Figure 2b shows how they are grouped together and related, forming a simple pattern of association.

Figure 2: This figure shows how four fundamental reasoning elements can be joined (linked) to form a pattern of association.

Figure 3: This figure shows how a part of one pattern of association can be used with other fundamental reasoning elements to create another pattern of association.

Figure 4: This figure shows how a simple pattern of association can be used as a basic building block (a resource) for forming a more complex pattern of association.

Figure 5: This figure illustrates how a person can hold three contradictory patterns of association (the velocity decreased, the velocity increased and the velocity stayed the same) simultaneously, if there is no motivation to choose one above the other at that point in time.

Figure 6: This figure is an illustration to indicate the general structure of patterns of association. Patterns of association consist of fundamental reasoning elements, indicated by yellow circles, and other patterns of association, illustrated by other shapes and colors. The different colors and shapes simply indicate more complex patterns of association.

Figure7: This figure illustrates how a person can choose to disassemble a pattern of association at level 2 (Fig. 7a), which causes many more complex patterns of association to disassemble (Fig. 7b), changing the complete structure of that person's patterns of association.